\documentclass[10pt, journal, onecolumn, twoside]{IEEEtran}
\ifCLASSINFOpdf
   \usepackage[pdftex]{graphicx}
\else
\fi
\usepackage{amsmath}
\usepackage{amsfonts}
\usepackage{bm}
\begin{document}
%
\title{Evaluation of Error Probability of Classification Based on the Analysis of the Bayes Code: Extension and Example}
%
%
%

\author{Shota~Saito,~\IEEEmembership{Member,~IEEE,}
        and~Toshiyasu~Matsushima,~\IEEEmembership{Member,~IEEE}
\thanks{
This work was supported in part by JSPS KAKENHI Grant Numbers 
JP17K06446, 
JP19K04914, 
and JP19K14989.
}
\thanks{Shota Saito is with the Faculty of Informatics, Gunma University, 4-2, Maebashi, Gunma, 371-8510, Japan (e-mail: shota.s@gunma-u.ac.jp); Toshiyasu Matsushima is with the Department of Applied Mathematics, Waseda University, 3-4-1 Okubo, Shinjuku-ku, Tokyo, 169-8555, Japan (e-mail: toshimat@waseda.jp).}
}

\maketitle

\begin{abstract}
Suppose that we have two training sequences generated by parametrized distributions $P_{\theta^*}$ and $P_{\xi^*}$, where $\theta^*$ and $\xi^*$ are unknown true parameters. Given training sequences, we study the problem of classifying whether a test sequence was generated according to $P_{\theta^*}$ or $P_{\xi^*}$. This problem can be thought of as a hypothesis testing problem and our aim is to analyze the weighted sum of type-I and type-II error probabilities. Utilizing the analysis of the codeword lengths of the Bayes code, our previous study derived more refined bounds on the error probability than known previously. However, our previous study had the following deficiencies: i) the prior distributions of $\theta$ and $\xi$ are the same; ii) the prior distributions of two hypotheses are uniform; iii) no numerical calculation at finite blocklength. This study solves these problems. We remove the restrictions i) and ii) and derive more general results than obtained previously. To deal with problem iii), we perform a numerical calculation for a concrete model.
\end{abstract}


%
\IEEEpeerreviewmaketitle

\newtheorem{theorem}{Theorem}
\newtheorem{prop}{Proposition}
\newtheorem{condi}{Condition}
\newtheorem{defi}{Definition}
\newtheorem{lem}{Lemma}
\newtheorem{cor}{Corollary}
\newtheorem{proof}{Proof}
\newtheorem{rem}{Remark}
\newcommand{\argmax}{\mathop{\rm arg~max}\limits}
\newcommand{\argmin}{\mathop{\rm arg~min}\limits}

\section{Introduction}
\subsection{Problem Setup} \label{BasicSetup}
Let 
$
{\bm y}_1 = y_{1,1}, \ldots, y_{1,N} \in {\cal A}^N
$
be a sequence called the 1st training sequence and 
$
{\bm y}_2 = y_{2,1}, \ldots, y_{2,N} \in {\cal A}^N
$
be a sequence called the 2nd training sequence.
We denote by ${\cal D} := \{ {\bm y}_1, {\bm y}_2 \}$.
For simplicity, we assume that ${\cal A}$ is a finite set, but our main results also hold for infinite set ${\cal A}$.

Suppose that $y_{1,1}, \ldots, y_{1,N}$ are drawn i.i.d. from a probability mass function $P(\cdot | \theta^*)$ and $y_{2,1}, \ldots, y_{2,N}$ are drawn i.i.d. from a probability mass function $P(\cdot | \xi^*)$.
We assume that parametric models $\{ P(\cdot | \theta) : \theta \in \Theta \subset \mathbb{R}^{d_{1}} \}$ and $\{ P(\cdot | \xi) : \xi \in \Xi \subset \mathbb{R}^{d_{2}} \}$ are known, but the true parameters $\theta^* \in \Theta$ and $\xi^* \in \Xi$ are {\it not} known, where $\Theta \subset \mathbb{R}^{d_{1}}$ is a $d_{1}$-dimensional parameter space and $\Xi \subset \mathbb{R}^{d_{2}}$ is a $d_{2}$-dimensional parameter space.
For notational convenience, we use $P_{\theta}(\cdot)$ and $P(\cdot | \theta)$ interchangeably (also, $P_{\xi}(\cdot)$ and $P(\cdot | \xi)$ interchangeably).

Let ${\bm x} = x_{1}, \ldots, x_{n} \in {\cal A}^n$ be a sequence called the test sequence.
Regarding the length of training sequences $N$ and the length of a test sequence $n$, we assume $N=\alpha n$ for some $\alpha > 0$.
Suppose that $x_{1}, \ldots, x_{n}$ are drawn i.i.d. from either $P(\cdot | \theta^*)$ or $P(\cdot | \xi^*)$, but we do not know whether they are from $P(\cdot | \theta^*)$ or $P(\cdot | \xi^*)$.
Then, the classification problem we consider in this paper is described as follows:

{\it Given a set of training sequences ${\cal D}=\{ {\bm y}_1, {\bm y}_2 \}$ and a test sequence ${\bm x}$, we attempt to classify a test sequence ${\bm x}$ whether it is generated according to $P(\cdot | \theta^*)$ or $P(\cdot | \xi^*)$.}

\subsection{Related Work}
The above problem can be thought of as a hypothesis testing problem and the probability of error has been investigated by several previous studies in information theory (see, e.g., \cite{Gutman}, \cite{Hsu}, \cite{Kelly}, \cite{Merhav}, \cite{Saito}, \cite{Zhou}, \cite{Ziv}).

Among these works, Merhav and Ziv \cite{Merhav} and our previous study \cite{Saito} have treated a Bayesian setting.
In these works, prior distributions of hypotheses and parameters (i.e., $\theta$ and $\xi$) were assumed and the weighted sum of type-I and type-II error probabilities was investigated.
Merhav and Ziv \cite{Merhav} used {\it the method of types} and investigated the {\it first-order term} of the probability of error for Markov sources as $n \to \infty$.
In a different way from \cite{Merhav}, our previous study \cite{Saito} investigated not only the first-order term but also {\it finer order terms} of the probability of error for i.i.d. sources.
We noticed the close relationship between the probability of error and the codeword lengths of the Bayes code (see, e.g., \cite{Clarke and Barron}, \cite[Chapter 7]{HanKobayashi}, \cite{Matsushima}) and utilized {\it the analysis of its codeword lengths} to derive the results.

\subsection{Contribution} \label{Contribution}
Our previous study \cite{Saito} described in Section I-B had some deficiencies.
First, we imposed the strong assumption that the prior distributions of $\theta$ and $\xi$ are the same.
Second, we imposed the strong assumption that the prior distributions of the two hypotheses are uniform.\footnote{Previous study \cite{Merhav} also imposed the assumptions that the prior distributions of $\theta$ and $\xi$ are the same and that the prior distributions of two hypotheses are uniform.}
Finally, we did not provide the results of numerical calculations regarding the upper and lower bounds at finite blocklength.

This paper solves these problems. 
More precisely, we remove the restricted assumptions in \cite{Saito} and derive more general results than those in \cite{Saito}. 
Further, we perform a numerical calculation for a concrete model and investigate the behavior of the upper and lower bounds.

\subsection{Organization}
The rest of this paper is organized as follows.
In Section \ref{Formulation}, we formulate a hypothesis testing problem and define its probability of error.
Next, as a preliminary, we describe the Bayes code in Section \ref{Pre}. 
Then, in Section \ref{Main}, we prove upper and lower bounds of the probability of error.
In Section \ref{Example}, we show the results of the numerical calculation.

\section{Formulation of Hypothesis Testing and Its Error Probability} \label{Formulation}
The problem described in Section \ref{BasicSetup} can be thought of as a hypothesis testing problem with the following two hypotheses:
\begin{itemize}
\item $H_1$: the 1st training sequence ${\bm y}_1$ and the test sequence ${\bm x}$ are generated according to the same distribution.
\item $H_2$: the 2nd training sequence ${\bm y}_2$ and the test sequence ${\bm x}$ are generated according to the same distribution.
\end{itemize}

A decision rule of this hypothesis testing problem is defined as follows:
\begin{defi}
A decision rule $\Lambda({\cal D})$, derived from a set of training sequences ${\cal D}=\{ {\bm y}_1, {\bm y}_2 \}$, is a partition of the space ${\cal A}^n$ into two disjoint regions $\Lambda_1 ({\cal D})$ and $\Lambda_2 ({\cal D})$ whose union equals ${\cal A}^n$, i.e., $\Lambda_1 ({\cal D}) \cap \Lambda_2 ({\cal D}) = \emptyset$, $\Lambda_1 ({\cal D}) \cup \Lambda_2 ({\cal D}) = {\cal A}^n$.
If ${\bm x} \in \Lambda_i ({\cal D})$, $H_i$ is accepted.
\end{defi}

We assume a prior distribution of a parameter $\theta$ (resp.\ $\xi$) and denote by $\mu(\theta)$ (resp.\ $\nu(\xi)$). 
As in \cite{Merhav}, we define the conditional error probability $P_{\Lambda}(e|{\cal D})$ associated with a decision rule $\Lambda = \Lambda({\cal D})$ as
\begin{align}
P_{\Lambda}(e|{\cal D}) := \sum_{i=1}^{2} \pi_i \sum_{{\bm x} \in \overline{\Lambda}_i ({\cal D})} P({\bm x} | {\cal D}, H_i), \label{ConProbErr}
\end{align}
where $\pi_i$ is a prior probability of hypothesis $H_i$, $\overline{\Lambda}_i ({\cal D})$ is the complement set of $\Lambda_i ({\cal D})$, and $P({\bm x} | {\cal D}, H_i)$ is a conditional probability mass function of ${\bm x}$ given ${\cal D}$ and the fact that the hypothesis $H_i$ is true.
This conditional probability mass function is calculated as\footnote{The detail of the calculation of (\ref{ConditionalProb1'}) and (\ref{ConditionalProb2'}) is shown in Appendix \ref{a1}.}
\begin{align}
P({\bm x} | {\cal D}, H_1)
&= \frac{\int_{\Theta} P({\bm x} | \theta) P({\bm y}_1 | \theta) \mu(\theta) d \theta}{\int_{\Theta} P({\bm y}_1 | \theta) \mu(\theta) d \theta} \label{ConditionalProb1'} \\
&=: P({\bm x} | {\bm y}_1, H_1), \label{ConditionalProb1} \\
P({\bm x} | {\cal D}, H_2) &= \frac{\int_{\Xi} P({\bm x} | \xi) P({\bm y}_2 | \xi) \nu(\xi) d \xi}{\int_{\Xi} P({\bm y}_2 | \xi) \nu(\xi) d \xi} \label{ConditionalProb2'} \\
&=: P({\bm x} | {\bm y}_2, H_2) \label{ConditionalProb2}
\end{align}
by using the Bayes theorem and imposing the following assumptions:\footnote{Similar assumptions were imposed in \cite{Merhav}.}
\begin{itemize}
\item[1)] The parameters $\theta$ and $\xi$ are independent, i.e., $P({\bm \theta}) = \mu(\theta) \nu(\xi)$, where ${\bm \theta} := (\theta, \xi) \in \Theta \times \Xi$.
\item[2)] $P({\cal D} | {\bm \theta}, H_i) = P({\cal D} | {\bm \theta}) = P({\bm y}_1 | \theta) P({\bm y}_2 | \xi)$ for $i \in \{1,2\}$.
\item[3)] $P({\bm x} | {\cal D}, {\bm \theta}, H_1) = P({\bm x} | \theta)$ and $P({\bm x} | {\cal D}, {\bm \theta}, H_2) = P({\bm x} | \xi)$.
\end{itemize}

By substituting (\ref{ConditionalProb1}) and (\ref{ConditionalProb2}) into (\ref{ConProbErr}), we have
\begin{align}
P_{\Lambda}(e|{\cal D}) = \sum_{i=1}^{2} \pi_i \sum_{{\bm x} \in \overline{\Lambda}_i ({\cal D})} P({\bm x} | {\bm y}_i, H_i) \label{PeLambda}
\end{align}
and the decision rule $\Lambda^* = \Lambda^*({\cal D}) = \{\Lambda_1^*({\cal D}), \Lambda_2^*({\cal D}) \}$ that minimizes (\ref{PeLambda}) is given by
\begin{align}
\hspace{-2mm} \Lambda_1^*({\cal D}) &= \{ {\bm x} \in {\cal A}^n : \pi_1 P({\bm x} | {\bm y}_1, H_1) \geq \pi_2 P({\bm x} | {\bm y}_2, H_2) \}, \label{Lambda1}\\
\hspace{-2mm} \Lambda_2^*({\cal D}) &= \{ {\bm x} \in {\cal A}^n : \pi_2 P({\bm x} | {\bm y}_2, H_2) \geq \pi_1 P({\bm x} | {\bm y}_1, H_1) \}, \label{Lambda2}
\end{align}
where ties are broken arbitrarily.
The conditional error probability associated with the decision rule $\Lambda^*$ is
\begin{align}
P_{\Lambda^*}(e|{\cal D}) = \sum_{i=1}^{2} \pi_i \sum_{{\bm x} \in \overline{\Lambda_i^*} ({\cal D})} P({\bm x} | {\bm y}_i, H_i). \label{PeLambda*}
\end{align}

The quantity (\ref{PeLambda*}) will be investigated in Section \ref{Main}.
To do so, the notion of the Bayes code plays an important role.
Hence, we review this in the next section.

\section{Preliminary: Bayes Code} \label{Pre}
Let ${\cal S}$ be a source alphabet and consider a source sequence $s^m = s_1 \ldots s_m \in {\cal S}^m$, where $s_1 \ldots s_m$ are drawn i.i.d. from a probability mass function $P(\cdot | \eta^{*})$.
Suppose that a class of parametrized distribution of a source $\{P(\cdot | \eta) : \eta \in {\cal E} \subset \mathbb{R}^d \}$ is known, but the $d$-dimensional true parameter $\eta^{*} \in {\cal E} \subset \mathbb{R}^d$ is unknown.
To perform a lossless compression of $s^m$ in this situation, we can use the Bayes code, which is one of the universal source codes (see, e.g., \cite{Clarke and Barron}, \cite[Chapter 7]{HanKobayashi}, \cite{Matsushima}).

The coding probability $P_{C}(\cdot)$ of the Bayes code is defined so that it minimizes the Bayes risk function defined as 
\begin{align}
\int_{{\cal E}} w(\eta) \sum_{s^m \in {\cal S}^m} P(s^m | \eta) \ln \frac{P(s^m | \eta)}{P_{C}(s^m)} d \eta, \label{Bayesrisk}
\end{align}
where $w(\eta)$ is a prior distribution of $\eta$.
In other words, the Bayes code is the optimal code in the sense that it minimizes the mean codeword length averaged with the prior $w(\eta)$.
The coding probability $P^\star_{C}(s^m)$ that minimizes (\ref{Bayesrisk}) is given by 
$P^\star_{C}(s^m) = \int_{{\cal E}} P(s^m | \eta) w(\eta) d \eta$ (see, e.g., \cite{Matsushima}), and the codeword length of the Bayes code $\ell_{\rm Bayes}(s^m)$ is given by
\begin{align}
\ell_{\rm Bayes}(s^m) 
= - \ln P^\star_{C}(s^m)
= - \ln \int_{{\cal E}}  P(s^m | \eta) w(\eta) d \eta. \label{CodewordLengthBayesCode}
\end{align}

In view of \cite{Clarke and Barron} and \cite[Chapter 7.8]{HanKobayashi}, we have the following lemma.
\begin{lem} \label{CB}
We assume the following conditions:
\begin{itemize}
\item[$1)$] The probability mass function $P(\cdot | \eta)$ is twice continuously differentiable at $\eta^{*}$, and there exists a $\delta > 0$ such that for $j, k \in \{1, 2, \ldots, d\}$, 
\begin{align}
\mathbb{E}_{P(s | \eta^*)} \left[ \sup_{\eta: \| \eta - \eta^* \| < \delta} \left( \frac{\partial^2}{\partial \eta_j \partial \eta_k} \ln P(S | \eta) \right )^2 \right] < \infty
\end{align}
and 
\begin{align}
\mathbb{E}_{P(s | \eta^*)} \left[\left( \left. \frac{\partial}{\partial \eta_j} \ln P(S | \eta) \right |_{\eta = \eta^*} \right)^2 \right] < \infty.
\end{align}
\item[$2)$] Let $D(P_{\eta^*} \| P_{\eta}) := \sum_{s} P(s | \eta^*) \ln \frac{P(s | \eta^*)}{P(s | \eta)}$ be the relative entropy and $J(\eta^*)$ be the $d \times d$ matrix defined by
\begin{align}
J(\eta^*) := \left[ \left. \frac{\partial^2}{\partial \eta_j \partial \eta_k} D(P_{\eta^*} \| P_{\eta}) \right|_{\eta = \eta^*} \right]_{j, k = 1, \ldots, d}.
\end{align}
Then, $D(P_{\eta^*} \| P_{\eta})$ is twice continuously differentiable at $\eta^*$, with $J(\eta^*)$ positive definite, and the prior $p(\eta)$ is continuous and positive at $\eta^*$.
\item[$3)$] For every open set ${\cal N}$ containing $\eta^*$ and every $\delta > 0$, 
\begin{align}
\mathbb{P} \left[w(\overline{{\cal N}} | S^m) > \delta \right] = o \left( \frac{1}{\ln m} \right),
\end{align}
where $\overline{{\cal N}}$ is the complement of ${\cal N}$ and $w(\cdot | S^m)$ is the posterior distribution of $\eta$ given $S^m$.
\end{itemize}

Under the conditions 1) -- 3) and a stationary memoryless source, we have
\begin{align}
\ell_{\rm Bayes}(s^m) = \ln \frac{1}{P(s^m | \hat{\eta})}  + \frac{d}{2} \ln \frac{m}{2 \pi} + \ln\frac{\sqrt{\det I(\hat{\eta})}}{w(\hat{\eta})} + o(1), \label{bayesL2}
\end{align}
as $m \to \infty$, where $\hat{\eta}=\hat{\eta}(s^m)$ is a maximum likelihood estimator and $I(\eta)$ is the $d \times d$ Fisher information matrix defined as
\begin{align}
I(\eta) := \mathbb{E}_{P(\cdot|\eta^*)}\left[ \frac{\partial}{\partial \eta_j} \ln P(S | \eta) \frac{\partial}{\partial \eta_k} \ln P(S | \eta) \right]_{j, k = 1, \ldots, d}.
\end{align}
\end{lem}

\begin{rem} \label{LengthandProbability}
Regarding the codeword length of the Bayes code, from (\ref{CodewordLengthBayesCode}) and (\ref{bayesL2}), we have
\begin{align}
\int_{{\cal E}} P(s^m | \eta) w(\eta) d \eta = P(s^m | \hat{\eta}) \left(\frac{m}{2 \pi} \right)^{-\frac{d}{2}} \left(\frac{\sqrt{\det I(\hat{\eta})}}{w(\hat{\eta})}\right)^{-1} e^{o(1)} \label{CodewordLengthBayesCodeII}
\end{align}
as $m \to \infty$.
On the other hand, regarding the term $P({\bm x} | {\bm y}_i, H_i)$ appeared in the error probability (\ref{PeLambda*}), the numerator and the denominator of the right-hand side of (\ref{ConditionalProb1'}) and (\ref{ConditionalProb2'}) take the similar form as in the left-hand side of (\ref{CodewordLengthBayesCodeII}).
This connection between $P({\bm x} | {\bm y}_i, H_i)$ and the codeword length of the Bayes code is one of the key points in the proofs of our main results.
\end{rem}

\section{Upper and Lower Bounds of $P_{\Lambda^*}(e|{\cal D})$} \label{Main}
We impose the following condition, which is similar to the condition in Lemma \ref{CB}.
\begin{condi} \label{MainCondition}
\begin{itemize}
\item[$1)$] The probability mass function $P(\cdot | \theta)$ and $P(\cdot | \xi)$ are twice continuously differentiable at $\theta^{*}$ and $\xi^*$ respectively, and there exist $\delta > 0$ and $\gamma > 0$ such that for $j, k \in \{1, 2, \ldots, d\}$, 
\begin{align}
\mathbb{E}_{P(x | \theta^*)} \left[ \sup_{\theta: \| \theta - \theta^* \| < \delta} \left( \frac{\partial^2}{\partial \theta_j \partial \theta_k} \ln P(X | \theta) \right )^2 \right] &< \infty, \\
\mathbb{E}_{P(x | \xi^*)} \left[ \sup_{\xi: \| \xi - \xi^* \| < \gamma} \left( \frac{\partial^2}{\partial \xi_j \partial \xi_k} \ln P(X | \xi) \right )^2 \right] &< \infty,
\end{align}
and 
\begin{align}
\mathbb{E}_{P(x | \theta^*)} \left[\left( \left. \frac{\partial}{\partial \theta_j} \ln P(X | \theta) \right |_{\theta = \theta^*} \right)^2 \right] &< \infty, \\
\mathbb{E}_{P(x | \xi^*)} \left[\left( \left. \frac{\partial}{\partial \xi_j} \ln P(X | \xi) \right |_{\xi = \xi^*} \right)^2 \right] &< \infty.
\end{align}
\item[$2)$] Let $J_1 (\theta^*)$ be the $d_1 \times d_1$ matrix and $J_2 (\xi^*)$ be the $d_2 \times d_2$ matrix defined by
\begin{align}
J_1 (\theta^*) &:= \left[ \left. \frac{\partial^2}{\partial \theta_j \partial \theta_k} D(P_{\theta^*} \| P_{\theta}) \right|_{\theta = \theta^*} \right]_{j, k = 1, \ldots, d_1}, \\
J_2 (\xi^*) &:= \left[ \left. \frac{\partial^2}{\partial \xi_j \partial \xi_k} D(P_{\xi^*} \| P_{\xi}) \right|_{\xi = \xi^*} \right]_{j, k = 1, \ldots, d_2}.
\end{align}
Then, $D(P_{\theta^*} \| P_{\theta})$ and $D(P_{\xi^*} \| P_{\xi})$ are twice continuously differentiable at $\theta^*$ and $\xi^*$ respectively, with $J_1 (\theta^*)$ and $J_2 (\xi^*)$ positive definite, and the prior $\mu(\theta)$ and $\nu(\xi)$ are continuous and positive at $\theta^*$ and $\xi^*$ respectively.
\item[$3)$] For every open set ${\cal N}_1$ containing $\theta^*$ and every open set ${\cal N}_2$ containing $\xi^*$, it holds that
\begin{align}
\mathbb{P} \left[\mu(\overline{{\cal N}_1} | X^n) > \delta \right] &= o \left( \frac{1}{\ln n} \right), \\
\mathbb{P} \left[\nu(\overline{{\cal N}_2} | X^n) > \gamma \right] &= o \left( \frac{1}{\ln n} \right)
\end{align}
for any $\delta > 0$ and $\gamma > 0$, where $\overline{{\cal N}_1}$ (resp.\ $\overline{{\cal N}_2}$) is the complement of ${\cal N}_1$ (resp.\ ${\cal N}_2$) and $\mu(\cdot | X^n)$ (resp.\ $\nu(\cdot | X^n)$) is the posterior distribution of $\theta$ given $X^n$ (resp.\ the posterior distribution of $\xi$ given $X^n$).
\end{itemize}
\end{condi}

The next theorem shows a lower bound of $-\ln P_{\Lambda^*}(e | {\cal D})$.
\begin{theorem} \label{MainBinaryLower}
Under Condition \ref{MainCondition}, it holds that
\begin{align}
- \ln P_{\Lambda^*}(e | {\cal D})  \geq n C(P_{\theta^*}, P_{\xi^*}) + \frac{\min \{d_1, d_2 \}}{2} \ln \left(1+\frac{1}{\alpha} \right) - \ln (\max \{\pi_1, \pi_2 \}) + o \left( 1 \right) \label{MainResultLower}
\end{align} 
almost surely as $n \to \infty$, where $C(P_{\theta^*}, P_{\xi^*})$ is the Chernoff information defined as
\begin{align}
C(P_{\theta^*}, P_{\xi^*}) := \sup_{\lambda \in (0,1)} - \ln \sum_{x \in {\cal A}} P(x | \theta^*)^\lambda P(x | \xi^*)^{1-\lambda}.
\end{align}
\end{theorem}

\begin{IEEEproof}
From (\ref{Lambda1}), (\ref{Lambda2}), and (\ref{PeLambda*}), we have  
\begin{align}
P_{\Lambda^*}(e|{\cal D}) 
&= \sum_{{\bm x} \in {\cal A}^n} \min \{ \pi_1 P({\bm x} | {\bm y}_1, H_1), \pi_2 P({\bm x} | {\bm y}_2, H_2) \} \notag \\
& \leq \pi_1^\lambda \pi_2^{1-\lambda} \sum_{{\bm x} \in {\cal A}^n} [ P({\bm x} | {\bm y}_1, H_1)]^{\lambda} [ P({\bm x} | {\bm y}_2, H_2)]^{1-\lambda}, \label{BoundofPeLambda*}
\end{align}
for all $\lambda \in (0,1)$, where the inequality is due to $\min \{\alpha, \beta \} \leq \alpha^{\lambda} \beta^{1-\lambda}$ for all $\alpha, \beta \in \mathbb{R}_+$.\footnote{$\mathbb{R}_+$ denotes the set of non-negative real numbers.}

Next, we evaluate $P({\bm x} | {\bm y}_i, H_i)$ ($i=1,2$) in the right-hand side of (\ref{BoundofPeLambda*}).
From Lemma \ref{CB}, we have
\begin{align}
P ({\bm x} | {\bm y}_1, H_1) &= \prod_{j=1}^{n} P(x_j | \theta^*) \left(\frac{1+\alpha}{\alpha} \right)^{-\frac{d_1}{2}} e^{o(1)},  \label{ConditionalProbII} \\
P ({\bm x} | {\bm y}_2, H_2) &= \prod_{j=1}^{n} P(x_j | \xi^*) \left(\frac{1+\alpha}{\alpha} \right)^{-\frac{d_2}{2}} e^{o(1)}  \label{ConditionalProbII'}
\end{align}
almost surely as $n \to \infty$ (see Remark \ref{LengthandProbability} regarding the connection between $P ({\bm x} | {\bm y}_i, H_i)$ and Lemma \ref{CB}; see Appendix \ref{a2} for the detail derivation of (\ref{ConditionalProbII}) and (\ref{ConditionalProbII'})).

Now, substituting (\ref{ConditionalProbII}) and (\ref{ConditionalProbII'}) into (\ref{BoundofPeLambda*}) and using the following equality
\begin{align}
\sum_{{\bm x} \in {\cal A}^n} \prod_{j=1}^{n} P(x_j | \theta^*)^{\lambda} P(x_j | \xi^*)^{1-\lambda} 
& =\prod_{l=1}^{n} \sum_{x_l \in {\cal A}}P(x_l | \theta^*)^{\lambda} P(x_l | \xi^*)^{1-\lambda} \\
& =\left(\sum_{x \in {\cal A}}P(x | \theta^*)^{\lambda} P(x | \xi^*)^{1-\lambda} \right)^n,
\end{align}
we obtain
\begin{align}
-\ln P_{\Lambda^*}(e|{\cal D}) \geq & n \left \{ - \ln \sum_{x \in {\cal A}} P(x | \theta^*)^\lambda P(x | \xi^*)^{1-\lambda} \right \} \notag \\
& + \left \{ \left( \frac{d_{1}}{2} - \frac{d_{2}}{2} \right) \lambda + \frac{d_{2}}{2}\right \} \ln \left( 1+ \frac{1}{\alpha} \right) \notag \\
& + \left \{ (\ln \pi_2 - \ln \pi_1) \lambda - \ln \pi_2 \right \} + o(1) \label{BoundofPeLambda*II}
\end{align}
almost surely as $n \to \infty$.

Noticing that $\lambda \in (0,1)$ is arbitrary, we take $\sup_{\lambda \in (0,1)}$ on the right-hand side of (\ref{BoundofPeLambda*II}).
We use the properties of supremum and infimum\footnote{More precisely, we use $\sup (A+B) \geq \sup A + \inf B$ twice from (\ref{BoundofPeLambda*II}) to (\ref{BoundofPeLambda*SupInf}).} and obtain
\begin{align}
-\ln P_{\Lambda^*}(e|{\cal D}) \geq & \sup_{\lambda \in (0,1)} \left [ n \left \{ - \ln \sum_{x \in {\cal A}} P(x | \theta^*)^\lambda P(x | \xi^*)^{1-\lambda} \right \} \right ] \notag \\
& + \inf_{\lambda \in (0,1)} \left [\left \{ \left( \frac{d_{1}}{2} - \frac{d_{2}}{2} \right) \lambda + \frac{d_{2}}{2}\right \} \ln \left( 1+ \frac{1}{\alpha} \right) \right ] \notag \\
& + \inf_{\lambda \in (0,1)} \left [ (\ln \pi_2 - \ln \pi_1) \lambda - \ln \pi_2 \right ] + o(1) \label{BoundofPeLambda*SupInf}
\end{align}
almost surely as $n \to \infty$.

The first term on the right-hand side of (\ref{BoundofPeLambda*SupInf}) is $n C(P_{\theta^*}, P_{\xi^*})$ due to the definition of the Chernoff information.
The second term on the right-hand side of (\ref{BoundofPeLambda*SupInf}) is calculated as
\begin{align}
\inf_{\lambda \in (0,1)} \left \{ \left( \frac{d_{1}}{2} - \frac{d_{2}}{2} \right) \lambda + \frac{d_{2}}{2}\right \}
= \begin{cases}
    \frac{d_{2}}{2} & (d_{1} \geq d_{2}), \\
    \frac{d_{1}}{2} & (d_{1} < d_{2}). 
  \end{cases}
\end{align}
Finally, the third term on the right-hand side of (\ref{BoundofPeLambda*SupInf}) is calculated as
\begin{align}
\inf_{\lambda \in (0,1)} \left [ (\ln \pi_2 - \ln \pi_1) \lambda - \ln \pi_2 \right ]
= \begin{cases}
    -\ln \pi_1 & (\pi_1 \geq \pi_2), \\
    -\ln \pi_2 & (\pi_1 < \pi_2).
  \end{cases}
\end{align}
Putting together the pieces, we obtain Theorem \ref{MainBinaryLower}.
\end{IEEEproof}

Next, we show an upper bound of $-\ln P_{\Lambda^*}(e | {\cal D})$.
\begin{theorem} \label{MainBinaryUpper}
Under Condition \ref{MainCondition}, we have
\begin{align}
-\ln P_{\Lambda^*}(e | {\cal D}) \leq n C (P_{\theta^*}, P_{\xi^*}) +\frac{1}{2} \ln \left(\frac{4n}{c^2} \right) + \frac{\max \{d_1, d_2 \}}{2} \ln \left(1+\frac{1}{\alpha} \right) - \ln (\min \{\pi_1, \pi_2 \}) + o \left( 1 \right) \label{MainResultUpper}
\end{align}
almost surely as $n \to \infty$, where the term $c$ is defined as in Appendix \ref{Defcn}.
\end{theorem}

As shown in  Appendix \ref{Defcn}, the term $c$ is a bit complicated.
Regarding this term, a more detailed discussion is given in Section \ref{Example}.

\begin{IEEEproof}
From (\ref{Lambda1}), (\ref{Lambda2}), and (\ref{PeLambda*}), we have
\begin{align}
\hspace{-1mm} P_{\Lambda^*}(e|{\cal D})= \sum_{{\bm x} \in {\cal A}^n} \min \{\pi_1 P({\bm x} | {\bm y}_1, H_1), \pi_2 P({\bm x} | {\bm y}_2, H_2) \}. \label{UpperPeLambda*}
\end{align}
Substituting (\ref{ConditionalProbII}) and (\ref{ConditionalProbII'}) into (\ref{UpperPeLambda*}) and dividing into cases, we obtain
\begin{align}
P_{\Lambda^*}(e|{\cal D}) & \geq \sum_{{\bm x} \in {\cal A}^n} \min \left\{\prod_{j=1}^{n} P(x_j | \theta^*), \prod_{j=1}^{n} P(x_j | \xi^*) \right\}  \times \begin{cases}
    \pi_2 \left(\frac{1+\alpha}{\alpha} \right)^{-\frac{d_1}{2}} e^{o(1)}  & (d_{1} \geq d_{2},~ \pi_1 \geq \pi_2), \\
    \pi_1 \left(\frac{1+\alpha}{\alpha} \right)^{-\frac{d_1}{2}} e^{o(1)}  & (d_{1} \geq d_{2},~ \pi_1 < \pi_2), \\
    \pi_2 \left(\frac{1+\alpha}{\alpha} \right)^{-\frac{d_2}{2}} e^{o(1)}  & (d_{1} < d_{2}, ~ \pi_1 \geq \pi_2), \\ 
    \pi_1 \left(\frac{1+\alpha}{\alpha} \right)^{-\frac{d_2}{2}} e^{o(1)}  & (d_{1} < d_{2}, ~ \pi_1 < \pi_2), 
  \end{cases} \label{UpperPeLambda*II}
\end{align}
almost surely as $n \to \infty$.

On the other hand, from \cite[Theorem 2.1]{ZhouandLi}, there exists $c$ defined as in Appendix \ref{Defcn} such that
\begin{align}
\sum_{{\bm x} \in {\cal A}^n} \min \left\{\prod_{j=1}^{n} P(x_j | \theta^*), \prod_{j=1}^{n} P(x_j | \xi^*) \right\} \geq \frac{c}{\sqrt{n}} \exp\{-n C(P_{\theta^*}, P_{\xi*})\} \label{BoundZhou}
\end{align}
for sufficiently large $n$.
More precisely regarding ``sufficiently large $n$'', from \cite[Theorem 2.1 and Remark 1]{ZhouandLi}, the inequality (\ref{BoundZhou}) holds for $n$ such that
\begin{align} 
\sqrt{n} \overline{\sigma} \lambda^* (1-\lambda^*) \geq {\sf C}', \label{ineqN}
\end{align}
where 
\begin{align} 
{\sf C}' := \max \{ 2,~ 2(0.56{\sf C})^{3/2} \exp( -\sqrt{2 \pi} {\sf C}) \},
\end{align}
and $\overline{\sigma}$, $\lambda^*$, and ${\sf C}$ are defined as in Appendix \ref{Defcn}.

Hence, combining (\ref{UpperPeLambda*II}) and (\ref{BoundZhou}), we obtain (\ref{MainResultUpper}).
\end{IEEEproof}

\begin{rem}
From Theorems \ref{MainBinaryLower} and \ref{MainBinaryUpper}, we see that the difference between upper and lower bounds of $(-1/n)\ln P_{\Lambda^*}(e | {\cal D})$ is $O(\ln (n) / n)$, which goes to zero as $n \to \infty$.
\end{rem}

\begin{rem}
Theorems \ref{MainBinaryLower} and \ref{MainBinaryUpper} show that the first-order term is the Chernoff information $C(P_{\theta^*}, P_{\xi^*})$, which is the best asymptotic achievable exponent of the weighted sum of type-I and type-II error probabilities for i.i.d. sources when two probability distributions are {\it known} (see, e.g., \cite[Chapter 11.9]{Cover}).
\end{rem}

\section{Numerical Calculation} \label{Example}
In this section, we perform a numerical calculation for a simple model to check the behavior of the upper and lower bounds in Theorems \ref{MainBinaryLower} and \ref{MainBinaryUpper} at finite blocklength.
Suppose that the 1st training sequence $y_{1,1}, \ldots, y_{1,N}$ are drawn i.i.d. from a probability mass function
\begin{align}
P(y | \theta^*) = \begin{cases}
    \theta^* & (y=1), \\
    1- \theta^* & (y=0), \\
    0 & ({\rm otherwise}),
  \end{cases}
\end{align}
where $\theta^* \in (0, 1)$ is a one-dimensional parameter.
Also, suppose that the 2nd training sequence $y_{2,1}, \ldots, y_{2,N}$ are drawn i.i.d. from a probability mass function
\begin{align}
P(y | \xi^*) = \begin{cases}
    \xi^* & (y=1), \\
    1- \xi^* & (y=0), \\
    0 & ({\rm otherwise}),
  \end{cases}
\end{align}
where $\xi^* \in (0, 1)$ is a one-dimensional parameter.

In this case, the Chernoff information is calculated as 
\begin{align} 
C (P_{\theta^*}, P_{\xi^*}) = -\ln[(\theta^*)^{1-\lambda^*} (\xi^*)^{\lambda^*} + (1-\theta^*)^{1-\lambda^*} (1-\xi^*)^{\lambda^*}],
\end{align}
where $\lambda^*$ is given by (\ref{deflam*}).

From \cite[Remark 1 and Section 2.3]{ZhouandLi}, the term $c$ in the right-hand side of (\ref{MainResultUpper}) is given by (\ref{defcn}) with
\begin{align} 
{\sf C} &:= \left | \ln \frac{\theta^* (1-\xi^*)}{\xi^* (1-\theta^*)} \right|, \\
\overline{\sigma} &:= \sqrt{\left \{ \ln \frac{\theta^* (1-\xi^*)}{\xi^* (1-\theta^*)} \right \}^2 p_{\lambda^*} (1-p_{\lambda^*})}, \\
p_{\lambda^*} &:= \frac{(\theta^*)^{1-\lambda^*} (\xi^*)^{\lambda^*}}{(\theta^*)^{1-\lambda^*} (\xi^*)^{\lambda^*} + (1-\theta^*)^{1-\lambda^*} (1-\xi^*)^{\lambda^*}}.
\end{align}

We set $\alpha = 2$, $\pi_1 = \pi_2 = 1/2$, and consider the following two cases:
\begin{itemize}
\item Case 1: $\theta^* = 0.55$, $\xi^* = 0.45$.
\item Case 2: $\theta^* = 0.3$, $\xi^* = 0.7$.
\end{itemize}
From simple calculation, we have $\lambda^* = 1/2$ in Cases 1 and 2.

From (\ref{ineqN}), the inequality (\ref{MainResultUpper}) holds for $n$ such that $n \geq 1590$ in Case 1 and $n \geq 90$ in Case 2.
Thus, we plot the upper bound (\ref{MainResultUpper}) and lower bound (\ref{MainResultLower}) in the range $1590 \leq n \leq 3000$ in Case 1 and $90 \leq n \leq 3000$ in Case 2.

\begin{figure} [t] 
\centering
\includegraphics[width=9.5cm,clip]{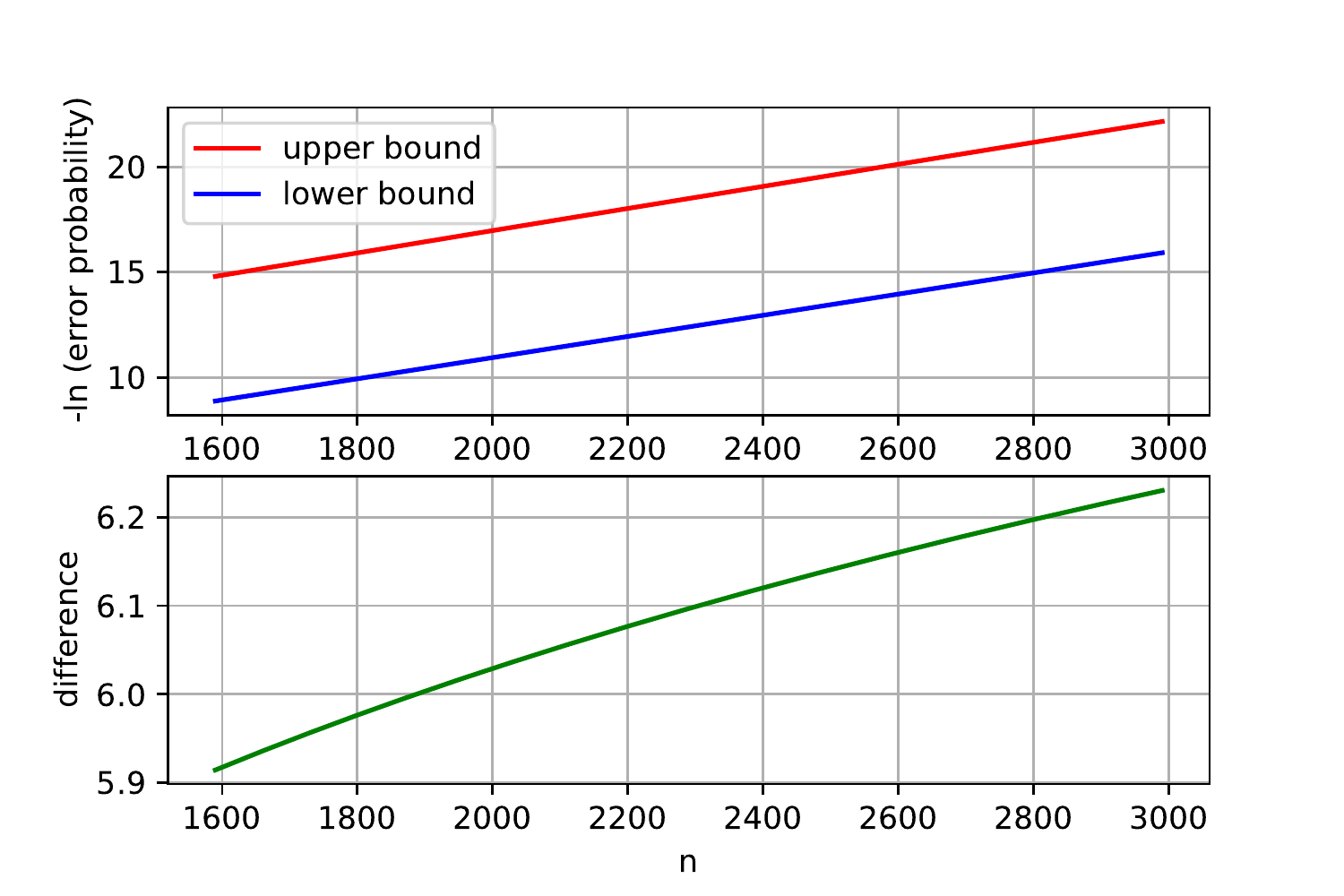}
\caption{The graph of Case 1 ($\theta^* = 0.55$, $\xi^* = 0.45$). The upper graph shows the bounds of $-\ln P_{\Lambda^*}(e | {\cal D})$; the red line is the upper bound (\ref{MainResultUpper}) and the blue line is the lower bound (\ref{MainResultLower}). The lower graph shows the difference between the upper and lower bounds.}
\label{fig1}
\end{figure}

\begin{figure} [t] 
\centering
\includegraphics[width=9.5cm,clip]{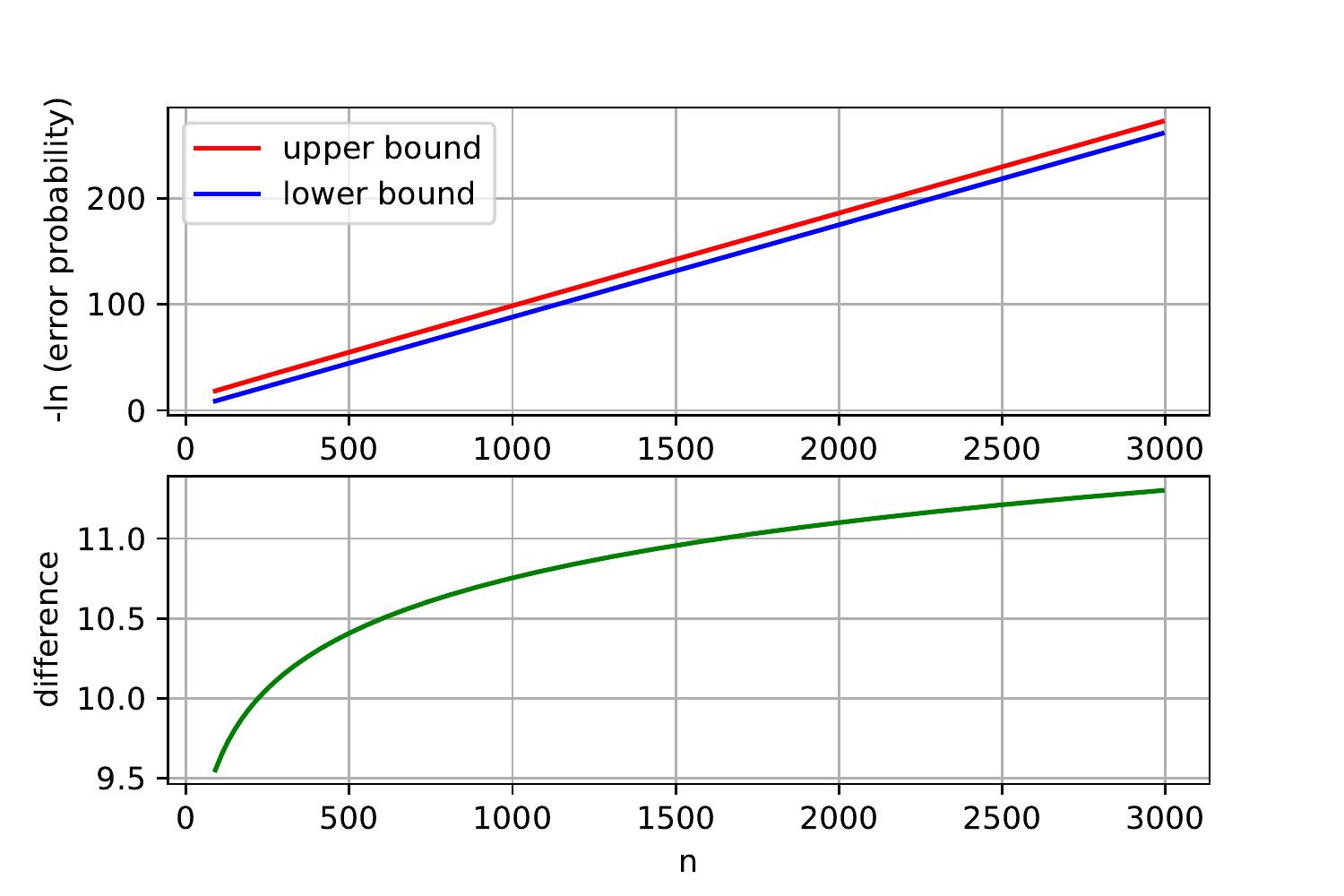}
\caption{The graph of Case 2 ($\theta^* = 0.3$, $\xi^* = 0.7$). The upper graph shows the bounds of $-\ln P_{\Lambda^*}(e | {\cal D})$; the red line is the upper bound (\ref{MainResultUpper}) and the blue line is the lower bound (\ref{MainResultLower}). The lower graph shows the difference between the upper and lower bounds.}
\label{fig2}
\end{figure}

The results of the numerical calculation in Case 1 are shown in Figure \ref{fig1}.
The upper graph in Figure \ref{fig1} shows the upper bound (\ref{MainResultUpper}) and lower bound (\ref{MainResultLower}) of $-\ln P_{\Lambda^*}(e | {\cal D})$.
Moreover, the lower graph in Figure \ref{fig1} shows the difference between the upper and lower bounds.
Similarly, the results in Case 2 is shown in Figure \ref{fig2}.

Comparing upper and lower bounds of $-\ln P_{\Lambda^*}(e | {\cal D})$ in Cases 1 and 2, the values in Case 1 are smaller than those in Case 2.
That is, the value of $P_{\Lambda^*}(e | {\cal D})$ is larger in Case 1 than in Case 2.
This can be explained by the fact that the more similar values of $\theta^*$ and $\xi^*$ make the hypothesis testing problem more difficult.

\appendices
\section{Proof of (\ref{ConditionalProb1'}) and (\ref{ConditionalProb2'})} \label{a1}
We show only (\ref{ConditionalProb1'}) because we can show (\ref{ConditionalProb2'}) in the same way.
From the assumptions 1) -- 3) and the Bayes theorem, we have
\begin{align}
&P({\bm x} | {\cal D}, H_1)
= \frac{P({\bm x},{\cal D} | H_1)}{P({\cal D} | H_1)} 
= \frac{P({\bm x}, {\cal D} | H_1)}{P({\cal D})} \\
&= \frac{\int_{\Theta} \int_{\Xi} P({\bm x}, {\cal D} | \theta, \xi, H_1) \mu(\theta) \nu(\xi) d\theta d\xi}{\int_{\Theta} \int_{\Xi} P({\cal D}| \theta, \xi) \mu(\theta) \nu(\xi) d\theta d\xi} \\
&= \frac{\int_{\Theta} \int_{\Xi} P({\bm x}|{\cal D}, \theta, \xi, H_1) P({\cal D} |\theta, \xi, H_1) \mu(\theta) \nu(\xi) d\theta d\xi}{\int_{\Theta} \int_{\Xi} P({\cal D}| \theta, \xi) \mu(\theta) \nu(\xi) d\theta d\xi} \\
&= \frac{\int_{\Theta} \int_{\Xi} P({\bm x} | \theta) P({\bm y}_1| \theta)P({\bm y}_2| \xi) \mu(\theta) \nu(\xi) d\theta d\xi}{\int_{\Theta}\int_{\Xi} P({\bm y}_1| \theta)P({\bm y}_2| \xi) \mu(\theta) \nu(\xi) d\theta d\xi} \\
&= \frac{\int_{\Theta}P({\bm x} | \theta) P({\bm y}_1| \theta) \mu(\theta) d\theta \int_{\Xi} P({\bm y}_2| \xi) \nu(\xi) d\xi}{\int_{\Theta} P({\bm y}_1| \theta) \mu(\theta) d\theta \int_{\Xi} P({\bm y}_2| \xi) \nu(\xi) d\xi} \\
&= \frac{\int_{\Theta}P({\bm x} | \theta) P({\bm y}_1| \theta) \mu(\theta) d\theta}{\int_{\Theta} P({\bm y}_1| \theta) \mu(\theta) d\theta}.
\end{align}

\section{Proof of (\ref{ConditionalProbII}) and (\ref{ConditionalProbII'})} \label{a2}
We show only (\ref{ConditionalProbII}) because we can show (\ref{ConditionalProbII'}) in the same way.
From Lemma \ref{CB}, we have
\begin{align}
\int_{\Theta} P({\bm y}_1 | \theta) \mu(\theta) d \theta 
= P({\bm y}_1 | \hat{\theta}) \left(\frac{N}{2 \pi} \right)^{-\frac{d_1}{2}} \left(\frac{\sqrt{\det I(\hat{\theta})}}{\mu(\hat{\theta})}\right)^{-1} e^{o(1)},  \label{Py}
\end{align}
where $\hat{\theta} = \hat{\theta}({\bm y}_1)$ is a maximum likelihood estimator.

Next, let ${\bm x} \circ {\bm y}_1$ denote a concatenation of a sequence ${\bm x}$ and a sequence ${\bm y}_1$.
Then, we have
\begin{align}
&\int_{\Theta}  P({\bm x} | \theta) P({\bm y}_1 | \theta) \mu(\theta) d \theta
\overset{(a)}{=}\int_{\Theta}  P({\bm x} \circ {\bm y}_1 | \theta) \mu(\theta) d \theta \\
&\overset{(b)}{=} P({\bm x} \circ {\bm y}_1 | \hat{\theta}') \left(\frac{n+N}{2 \pi} \right)^{-\frac{d_1}{2}} \left(\frac{\sqrt{\det I(\hat{\theta}')}}{\mu(\hat{\theta}')} \right)^{-1} e^{o(1)}, \label{Pxy}
\end{align}
where 
$(a)$ follows from the fact that $x_{1}, \ldots, x_{n} \sim {\rm i.i.d.}~P(\cdot | \theta)$ and $y_{1,1}, \ldots, y_{1,N} \sim {\rm i.i.d.}~P(\cdot | \theta)$, and $(b)$ is due to Lemma \ref{CB}, where $\hat{\theta}' = \hat{\theta}'({\bm x} \circ {\bm y}_1)$ is a maximum likelihood estimator.

Substituting (\ref{Py}) and (\ref{Pxy}) into (\ref{ConditionalProb1'}) and recalling that $N=\alpha n$, we obtain
\begin{align}
P ({\bm x} | {\bm y}_1, H_1) 
= \frac{P({\bm x} \circ {\bm y}_1 | \hat{\theta}')}{P({\bm y}_1 | \hat{\theta})} \left(\frac{1+\alpha}{\alpha} \right)^{-\frac{d_1}{2}} \frac{\mu(\hat{\theta}')}{\mu(\hat{\theta})} \frac{\sqrt{\det I(\hat{\theta})}}{\sqrt{\det I(\hat{\theta}')}} e^{o(1)}.
\end{align}
Noticing that $\hat{\theta} \to \theta^*$ and $\hat{\theta}' \to \theta^*$ almost surely as $n \to \infty$ and the continuity of $I(\cdot)$ and $\mu(\cdot)$ (due to the assumptions of the theorem), we have
\begin{align}
P ({\bm x} | {\bm y}_1, H_1) = \frac{P({\bm x} \circ {\bm y}_1 | \theta^*)}{P({\bm y}_1 | \theta^*)} \left(\frac{1+\alpha}{\alpha} \right)^{-\frac{d_1}{2}} e^{o(1)}
\end{align}
almost surely as $n \to \infty$.
Finally, since 
\begin{align}
\frac{P({\bm x} \circ {\bm y}_1 | \theta^*)}{P({\bm y}_1 | \theta^*)} 
= \frac{P({\bm x} | \theta^*) P({\bm y}_1 | \theta^*)}{P({\bm y}_1 | \theta^*)} 
= P({\bm x} | \theta^*)
=\prod_{j=1}^{n} P(x_j | \theta^*),
\end{align}
we obtain (\ref{ConditionalProbII}).

\section{Definition of $c$} \label{Defcn}
Let
\begin{align} 
\lambda^* := \argmax_{\lambda \in (0,1)} \left [ - \ln \sum_{x \in {\cal A}} P(x | \theta^*)^\lambda P(x | \xi^*)^{1-\lambda} \right] \label{deflam*}
\end{align}
and we define the probability mass function $\varphi_{\lambda^*}(a)$ for $a \in {\cal A}$ as
\begin{align} 
\varphi_{\lambda^*}(a) := \frac{P(a | \theta^*)^{\lambda^*} P(a | \xi^*)^{1-\lambda^*}}{\sum_{a' \in {\cal A}}P(a' | \theta^*)^{\lambda^*} P(a' | \xi^*)^{1-\lambda^*}}.
\end{align}
Then, we define $Z$ as
\begin{align} 
Z := \ln \frac{P(A | \theta^*)}{P(A | \xi^*)},
\end{align}
where $A$ is a random variable whose distribution is $\varphi_{\lambda^*}$, and define
\begin{align} 
\overline{\sigma} := \left( {\rm Var}[Z]\right)^{1/2}.
\end{align}
Further, let ${\sf C}$ be a positive constant satisfying
\begin{align} 
\mathbb{E}[|Z|^3] \leq {\sf C} \overline{\sigma}^2.
\end{align}

Under these notations, the term $c$ is defined as
\begin{align} 
c := \frac{\exp(-1.12 \sqrt{2 \pi} {\sf C})}{30 \overline{\sigma} \lambda^* (1-\lambda^*)}. \label{defcn}
\end{align}

\end{document}